\documentclass{aa}
\usepackage{graphicx}
\usepackage{subfig}
\usepackage{txfonts}
\usepackage{soul} 
\usepackage{xcolor}

\newcommand{\ac}[1]{{\color{blue}{#1}}}

\begin{document}

   \title{Modulation depth of the gyrosynchrotron  emission as identifier of fundamental sausage modes}


   \author{M. C\'ecere\inst{1,2}
   \and
          A. Costa\inst{1}
          \and
          T. Van Doorsselaere\inst{3}
          }

   \institute{Instituto de Astronom\'{\i}a Te\'orica y Experimental, CONICET-UNC, C\'ordoba, Argentina.
             \and
             Observatorio Astron\'omico de C\'ordoba, UNC, C\'ordoba, Argentina.
        \and
             Centre for mathematical Plasma Astrophysics, Department of Mathematics, KU Leuven, Celestijnenlaan 200B bus 2400, B-3000, Leuven, Belgium.
        }

   \date{}


  \abstract
  {}
  { We study the intensity, the modulation depth and the mean modulation depth of the  gyrosynchrotron (GS) radiation  as a function of the frequency and the line of sight (LOS) in fast sausage modes.}
   {By solving the 2.5D magnetohydrodynamics (MHD) ideal equations of a straight coronal loop considering the chromosphere and  with typical flaring plasma parameters  we analyse the  wavelet transform of the density and the GS emission for different radio frequencies and different spatial resolutions,  given impulsive and general  perturbations with energies in the microflare range.}
   {A wavelet analysis performed over the GS radiation emission  showed that a fast fundamental sausage mode of $\sim7\,\text{s}$ with a first harmonic  mode of $3\,\text{s}$ developed, for all the initial energy perturbations used.
   For both the high spatial resolution (central pixel integration) and the low spatial resolution (entire loop integration), the larger the radio frequency, the larger the modulation depth. However, high and low resolution integrations differ in that, the larger the LOS angle with respect to the loop axis, results in a larger and smaller modulation depth, respectively.
   }
   {Fast MHD modes triggered by instantaneous energy depositions of the order of a microflare energy are able to reproduce deep intensity modulation depths in radio emission as observed  in solar events.
   As the trends of the GS emission obtained by Reznikova, Antolin, and Van Doorsselaere (2014),  for a linear and forced oscillation, remain present when analysing a more general context, considering the chromosphere and where the sausage mode is triggered by a impulsive, nonlinear perturbation, it seems that the behaviour found can be used as observational identifiers of the presence of sausage modes with respect to other quasi-periodic pulsation features.
   It can be inferred from this that finite-amplitude sausage modes have the potential to generate the observed deep modulation depths.   }


   \maketitle
%

\section{Introduction}

Solar flares and stellar flares of M red dwarfs and solar-type active stars, which are still needed to last from several seconds to several hours,  are the highest-energy processes of their atmospheres, capable of releasing a large amount of energy
\citep{2021SSRv..217...66Z, 2020STP.....6a...3K}.  The stellar flare energies are comparable or often considerably higher than the most powerful solar flare detected \citep{1859MNRAS..20...13C} with a total energy of $\sim 10^{32}$ erg. The so-called superflares are flares with a total energy of over $10^{33}$ erg, however detected  in only $\sim{0.1}$ of all the observed stars \citep{2021ApJ...923L..33K}. The study of stellar flares and superflares has become an  important issue to investigate
whether the Sun is capable of producing a damaging
solar superflare.

In most of solar and stellar flares, quasi-periodic pulsations (QPPs) were detected in all spectral bands, from low-frequency radio waves
to high gamma-ray energies (for a review see e.g. \cite{2020STP.....6a...3K}  and reference therein). The observed QPP periods vary from
fractions of a second to tens of minutes and their duration is not longer than a few tens of oscillations \citep{2019PPCF...61a4024N}.

Several empirical similarities between stellar and  solar QPPs in flares \citep{2016ApJ...830..110C} strongly suggest that a common  underlying physical mechanism could explain  the phenomenon. Thus,  these  transient oscillations
seem to be   unique seismological tools to give account
of  non-spatially-resolved   plasma conditions and processes in stars.

While the physical mechanisms responsible for producing this omnipresent  behaviour are still not fully understood,  three main   mechanisms were proposed. The QPP signal
could be produced by   repetitive magnetic reconnections induced periodically by an magnetohydrodynamic (MHD) oscillation, the reconnection process itself could be repetitive and periodically  triggered internally in a spontaneous wave (self-oscillations)  or   QPP signals would be the emission of the local plasma modulated  by MHD oscillations \citep{2018SSRv..214...45M}.

Due to the similarity between QPP periods in most of  solar and stellar flares with respect to typical coronal MHD fast wave periods, we here investigate the emission properties of fast sausage modes as producers of  QPP signals.

Sausage  modes  manifest observationally as quasi-periodic pulsation patterns.
They are  produced by collective plasma perturbations in relatively thick and short magnetic structures  observed in flaring regions of low $\beta$ plasma and are useful tools to deduce plasma parameters difficult to measure directly, see e.g. \citet{2005LRSP....2....3N,2011ApJ...740...90V,2016SoPh..291.3139N,2016SoPh..291.3143V,2020ARA&A..58..441N,2020STP.....6a...3K}. The fast sausage mode is an essentially axisymmetric compressive mode where the magnetic field and the density perturbations are  almost perpendicular and the plasma motion is mainly perpendicular to the unperturbed loop axis.
Due to the expected short periods of the mode (of a few seconds), instruments of high time resolution observing at microwave and hard X-ray wavelengths are required \citep{2005A&A...439..727M,2008A&A...487.1147I}. Particularly, the first fundamental (global trapped) sausage mode observation, with a peak at $17~\text{s}$, was detected by \citet{2003A&A...412L...7N} using the Nobeyama Radioheliograph at $17~\text{GHz}$ and $34~\text{GHz}$.

Observations show that QPP that are detected in
the microwave and X-ray bands, have much deeper modulations depths than EUV-visible-light-MHD waves (sometimes up to  a hundred percent). Thus, it is important to provide a criterion to establish if the nature  of these two types of oscillations are  of the same nature   \citep{2006A&A...452..343N, 2020STP.....6a...3K}.

\cite{2012ApJ...748..140M,2014ApJ...785...86R,2015SoPh..290.1173K} modelled the modulation depth (hereafter MD) of the gyrosynchrotron (GS) emission by MHD oscillations using a specific geometrical perturbation of a given  monochromatic frequency and usual flaring parameters. In particular, \citet[hereafter R14]{2014ApJ...785...86R} found that the resulting MD of fast sausage modes is spatially dependent on the inhomogeneity of its oscillating  source, i.e., depending on the viewing angle; and the spatial resolution of the model used for the analysis.


More recently, \citet[hereafter C20]{2020A&A...644A.106C}  analysed  the capability of different types of perturbations associated with energy fluctuations of the solar corona to excite slow and fast sausage modes in solar flaring loops.
In flaring regions, the characteristic conductive heating time is much smaller than the radiative cooling one. Thus, typical impulsive disturbance are capable  of impulsively modify the entire internal temperature of the loop exciting a dominant fast sausage mode.

In this paper the aim is to analyse the  R14 results --specifically the importance of the viewing angle, the spatial resolution (a part of the loop or the entire loop) and the frequency variations in the MD of the GS emission-- considering the chromosphere and using a  global impulsive perturbation of overlapping frequencies with the capability to excite a sausage mode as found in C20. The purpose is to investigate if the R14 results obtained by a  linear and monochromatic analysis are also valid when perturbing in a more natural context, i.e., with a more general perturbation of energy, of the order of a microflare energy.
Also, to consider different  spatial resolutions is of interest to analyse the oscillatory features that could remain when the loop radiation is integrated as a whole in order to compare with different observing frames, e.g., the QPP description in a point-like stellar flaring loop.

In Section~\ref{sec:model} we detail the  model used to simulate the sausage mode oscillations and the tools  to emulate its GS emission. Results are shown in Section~\ref{sec:results}. Through a wavelet analysis on the density and the intensity parameters we study the dominant mode developed considering both a small central region of the loop and the entire loop. Also, in this section we show the GS intensity  and its relation with the magnetic field and the density. We also analyse the behaviour of the MD when the frequencies and the line of sight (LOS) viewing angles are changed. Finally, in Section~\ref{sec:conc}  we present the main conclusions.

\section{Model}\label{sec:model}

\subsection{Initial setup}
We solve the 2.5D ideal MHD equations to study the sausage oscillations of coronal loops based on the model proposed by C20.
Considering a  straight coronal loop of  typical flaring parameters in equilibrium with its surrounding medium (corona and chromosphere), we set the C20 geometrical dimensions (length $L=25\text{Mm}$ and width $w=6\text{Mm}$). We perturb the initial equilibrium configuration with a global energy deposition, that is, we instantaneously increase the temperature along the entire loop. This type of perturbation occurs in flaring regions where the conductive heating times are very small. In this scenario an additional heat applied to one end of the loop immediately heats up the entire loop. To analyse how this perturbation is able to produce GS emission, we choose as reference conditions the ``Base model'' initial plasma parameters of R14 (see the Corona and Loop columns of Table~\ref{tab:tablerez}).

We excite the initial states instantaneously increasing  the temperature of the reference loop: $\Delta T = (1, 3, 10, 20)~\text{MK}$ and we name each disturbed case as B1, B3, B10, and B20, respectively (see last four columns of Table~\ref{tab:tablerez}).   Each increment resembles the action of a microflare and is equivalent to an increment in the internal energy $\Delta E_i = \frac{n k  \Delta T}{\gamma -1} V = (1.45 ~10^{27}, 4.37~10^{27}, 1.45~10^{28},  2.91~10^{28})~ \text{erg}$, respectively, and where $V$ is the volume of the cylinder. The $\beta$-parameters, the sound and Alfvén speeds are also shown in the Table.

\begin{table*}[h]
 \centering
 \begin{tabular}{|c|c|c|c|c|c|c|}
  \hline
  & Corona     & Loop & B1  & B3 & B10 & B20      \\
  \hline
  \hline
  $n$ [cm$^{-3}$] & $2~10^{8}$ & $10^{10}$ & $10^{10}$& $10^{10}$& $10^{10}$& $10^{10}$   \\
  \hline
  $T$ [MK] & $2$ & $10$ & $11$ & $13$ & $20$ & $30$ \\
  \hline
  $B_z$  [G] & $56$ & $53$ & $53$& $53$& $53$& $53$ \\
  \hline
  $\beta$ & $0.0004$ & $0.13$ & $0.14$ & $0.17$ & $0.26$ & $0.37$  \\
  \hline
  $\Delta E_i~ [10^{27} \text{erg}]$ &  &  & $1.45$ & $4.37$ & $14.5$ & $29.1$  \\
  \hline
  $c_s$ [km/s] & $210$ & $470$ & $500$ & $540$ & $670$ & $940$ \\
  \hline
  $v_A$ [km/s] & $11000$ & $1470$ & $1470$& $1470$& $1470$& $1470$  \\

  \hline
 \end{tabular}
 \caption{System initial values for the different  simulated cases.}
 \label{tab:tablerez}
\end{table*}

\subsection{Radio emission}

We analyse the sausage radio emission  in different scenarios  treating the numerical data with  the FoMo-GS tool  \citep{2016FrASS...3....4V}. To describe the non-thermal electron distribution, we consider a distribution function $G(E,\mu)=u(E)g(\mu)$ where $u(E)$ is the energy and $g(\mu)$ is the angular distribution functions, respectively \citep{2010ApJ...721.1127F,2021ApJ...922..103K}. We take into account two electron distributions. A single power-law distribution with $u(E)=a~E^{-\delta}$ for $E_{min}<E<E_{max}$, where the electron spectral index $\delta=3.5$, $E_{min}=0.1~\text{MeV}$, and $E_{max}=10~\text{MeV}$. A non-zero non-thermal electron number density is used inside the loop: $n_{n-t}=n/200$.
For the double power-law distribution, we choose $\delta=1.5$ for $E_{min}<E\leq E_{break}$, and $\delta=3$ for $E_{break}\leq E<E_{max}$, where $E_{min}=0.01~\text{MeV}$, $E_{break}=0.5~\text{MeV}$, and $E_{max}=10~\text{MeV}$. For this distribution, the non-thermal electron number density inside the loop is $n_{n-t}=n/4000$. The distribution over the pitch angle is isotropic with $g(\mu)=0.5$.

\begin{figure}[h!]
    \centering
    \includegraphics[width=0.75\linewidth]{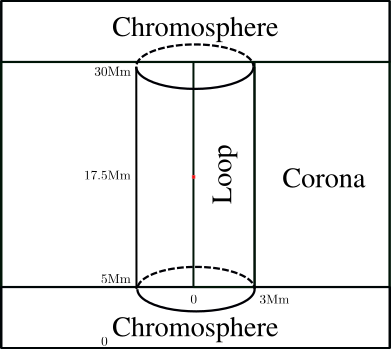}

    \caption{Scheme of the loop  and its environment. The small red point represents a pixel of $0\farcs5$. }
    \label{fig:squeme}
\end{figure}

 Given the axisymmetry of the problem, to calculate the GS emission of a cylindrical coronal loop, we have to construct 3 dimensional data from the 2.5D numerical result. Thus, we perform a revolution of the 2.5D data around the loop axis. We  consider two cases: 1) the integration along the LOS of a central pixel of $0\farcs5$ (equivalent to $5\times5$ simulation cells  around $(1,17.5)~\text{Mm}$, hereafter ``\textbf{Central pixel}'', shown in Fig.~\ref{fig:squeme}) and 2)  the integration of the whole loop, to resemble the signal obtained from distant point-like stars (hereafter ``\textbf{Entire loop}''). Our results do not show significant differences between the  single and the double power-law distribution of non-thermal electrons \citep{2022ApJ...937L..25S}. We here show the results of the  double-law spectrum parameters of R14.

\section{Results} \label{sec:results}

\subsection{Wavelet analysis}

\begin{figure}[h!]
    \centering

    \includegraphics[width=\linewidth]{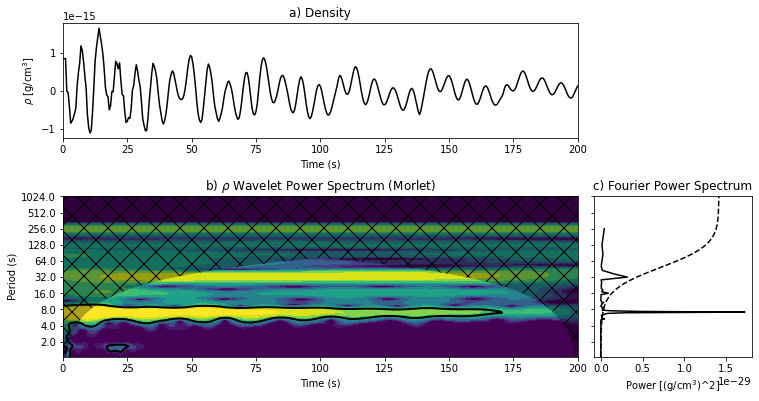}

    \caption{Density wavelet analysis. (a) Density along time measured at the {\bf Central pixel} of the B20 case. (b) Wavelet power spectrum. The black cross-lines show the part of the spectrum that is inside the cone of influence. The dominant period of $\sim7~\text{s}$ is enclosed by the black contours of $99\%$ confidence level.
    (c) Fourier power spectrum (solid line) and the $99\%$ confidence spectrum (dashed line).}
    \label{fig:wavelet_base}
\end{figure}

To analyse the modes excited  by the global impulsive perturbation in a  loop initially in equilibrium, we  wavelet analysed  the density of detrended data. We found that all cases present the same behaviour, except the amplitude of the Fourier power spectrum which increases an order of magnitude with  the increase of the  internal energy deposition.  Figure~\ref{fig:wavelet_base}  shows the wavelet analysis of the B20 case (the other cases's behaviours are similar and are not shown here). In panel (a) we show the density along time, measured at the {\bf Central pixel}. In panel (b) we see the wavelet power spectrum of the density during $200~\text{s}$. In panel (c) we see the Fourier power spectrum. A dominant period of $\sim7~\text{s}$ corresponding  to a fast sausage mode ($\tau\sim 2w/v_A$)  with an amplitude significantly high  during $~170~\text{s}$ is obtained.  A $\sim 32~\text{s}$ period of  a  slow mode ($\tau\sim L/c_s$) is also found, which is not significant.

\begin{figure}[h!]
    \centering
    \includegraphics[width=\linewidth]{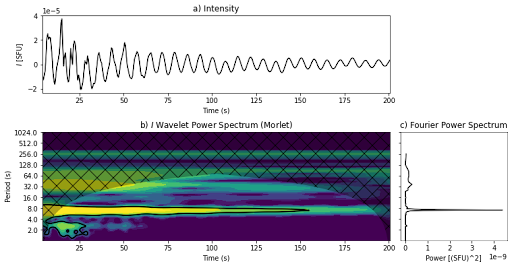}

    \caption{Intensity wavelet analysis. (a) Intensity along time measured at the \textbf{Central pixel} for B20 case, at $100~\text{GHz}$. Panels (b) and (c) describe the same features as in Fig.~\ref{fig:wavelet_base}.}
    \label{fig:wavelet_base_intensity}
\end{figure}

The Razin suppression (Razin 1960a, 1960b)\ac{\footnote{As the Razin effect states, the GS radiation is suppressed at frequencies where the refractive index becomes significantly less than unity.}} is negligible at frequencies larger than $1~\text{GHz}$, thus we analyse the cases at frequencies larger than $10~\text{GHz}$. We estimate the radiation emitted by the \textbf{Central pixel} and we calculate the GS emission at different frequencies ($10~\text{GHz}$ and $100~\text{GHz}$) for all cases. We only show here the B20 case ($100~\text{GHz}$) due to the similar behaviour between  the different frequencies except for the magnitude of the Fourier power spectrum. From Fig.~\ref{fig:wavelet_base_intensity}, we note that the fast sausage mode is the main mode observed. Also,  a first harmonic of $3~\text{s}$ has appeared, while the previous slow mode has disappeared. The sausage first harmonic corresponds to a higher frequency oscillation observed in the first $25\,\text{s}$.

\begin{figure}[h!]
    \centering
    \includegraphics[width=\linewidth]{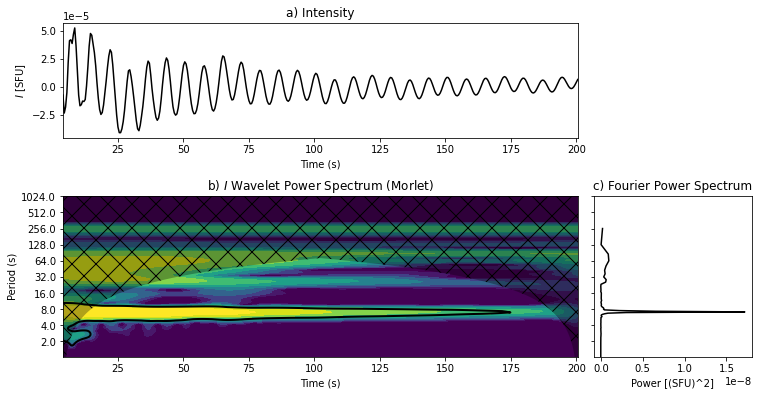}

    \caption{Same as Fig.~\ref{fig:wavelet_base_intensity} for the \textbf{Entire loop}.}
    \label{fig:wavelet_base_intensity_whole}
\end{figure}

We  see from Fig.~\ref{fig:wavelet_base_intensity_whole} that similar results are obtained if we consider the \textbf{Entire loop}. The higher frequencies appearing in Fig.~\ref{fig:wavelet_base_intensity}  (see e.g. panel (a))  in the first $25\,\text{s}$  are now nearly invisible due to the average performed.


\subsection{Imaging analysis}

\begin{figure}[h]
\centering
\begin{tabular}[b]{c@{}c@{}}
\includegraphics[width=.75\linewidth]
{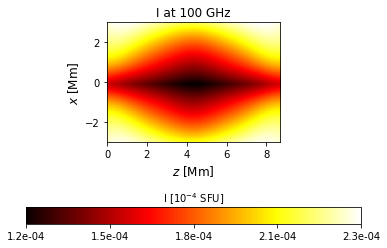} \\
\includegraphics[width=.75\linewidth]
{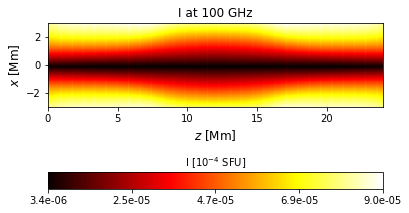}%
\end{tabular}%
\caption{Intensity obtained by integrating the radiation along the LOS  angle of $30^{\circ}$ (upper panel) and $85^{\circ}$ (lower panel) at frequency $100~\text{GHz}$, for case B20 at $t=10~\text{s}$.}
\label{fig.ivbasem}
\end{figure}

\begin{figure}[h]
\centering
\begin{tabular}[b]{@{}c@{}}
\includegraphics[width=.8\linewidth]{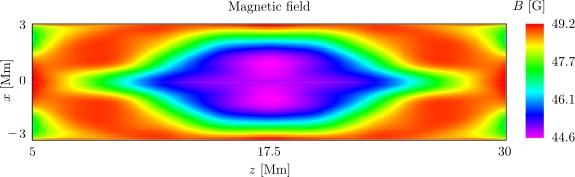} \\
\includegraphics[width=.8\linewidth]{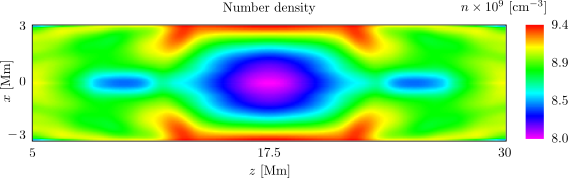}
\end{tabular}%
\caption{Magnetic field and number density at $t=10~\text{s}$ for B20.}
\label{fig.mag_dens}
\end{figure}

We  analyse the emission intensity by processing   the  numerical data (Fig.~\ref{fig.ivbasem}),  integrating the radiation along the LOS at the angle $30^{\circ}$ and $85^{\circ}$ and frequency $100~\text{GHz}$, for the B20 case at $10~\text{s}$. Note that when the LOS angle is smaller, the path length of the emission integration is larger and the intensity increases significantly, by at least an order of magnitude, when the angle varies from $85^{\circ}$ to $30^{\circ}$. The path length that comes from smaller angles necessarily crosses zones of maximum and minimum GS emission, therefore it is reasonable that the integrated emission obtained never reaches its minimum value as it does when the angle is transversal to the loop.
Also, a smooth modulation of the intensity which mainly responds to the density and the magnetic field pattern (see Fig.~\ref{fig.mag_dens}) is observed.
The lower intensity along the loop axis is related with the larger optical depth i.e., the integrated path is larger at the axis location.

\subsection{Radio frequencies analysis}

\begin{figure}[!h]
\centering
\begin{tabular}[b]{@{}c@{}}
\includegraphics[width=.5\linewidth]{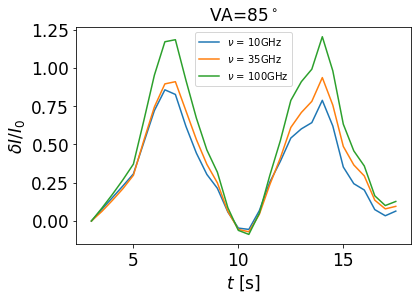}%
\includegraphics[width=.5\linewidth]{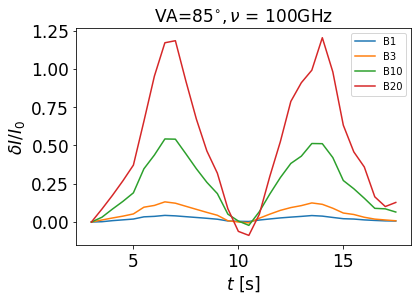}\\
\end{tabular}%
\caption{MD analysis for a given viewing angle. Left panel: MD measured along the \textbf{Central pixel} for B20 case at different frequencies. Right panel: Changes of MD through different cases calculated at $100~\text{GHz}$; both panels with a viewing angle of $85^{\circ}$.}
\label{fig.deltaIbasem_freqs}
\end{figure}

\begin{figure}[!h]
\centering
\begin{tabular}[b]{@{}c@{}}
\includegraphics[width=.5\linewidth]{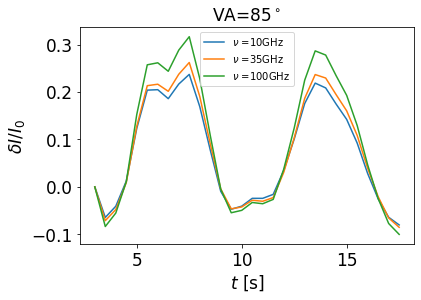}%
\includegraphics[width=.5\linewidth]{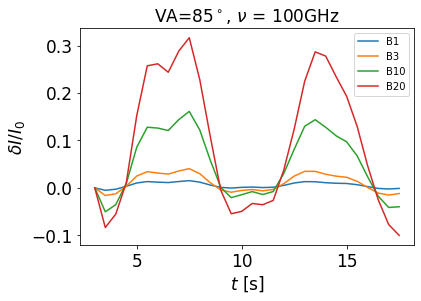}\\
\end{tabular}%
\caption{Same as Fig.~\ref{fig.deltaIbasem_freqs} for the \textbf{Entire loop}.}
\label{fig.deltaIbasem_whole_freqs}
\end{figure}

The left panel of Fig.~\ref{fig.deltaIbasem_freqs}  shows the MD ($\delta I/I_0 = (I_\nu(t)-I_\nu(0))/I_\nu(0)$,  where $I_\nu(t)$ is the time series of the intensity at the selected frequency $\nu$) for two oscillation periods  for radio frequencies $10~\text{GHz}$ (blue line), $35~\text{GHz}$ (orange line) and $100~\text{GHz}$ (green line), respectively and for B20 case, considering the \textbf{Central pixel} integration. When the loop is observed at $85^{\circ}$ we find slightly larger MDs with the increase of the frequency\footnote{The trend is not changed when the pixel location is changed.}.
The same occurs for the sequence of frequencies at angles $30^\circ$, $45^\circ$ and $60^\circ$ (not shown here).
This trend does not occur in R14 for certain angles e.g. $30^\circ$ and $45^\circ$.
This is due to  a particular pattern associated with the larger radio frequencies. The higher the frequency, the  smaller the  amplitude of the  oscillating intensity. However, as the intensity values are very small for large frequencies  ($I\propto \nu^{1.22-0.9\delta}$)   the denominator is small enough  to produce a significant  increase in the MD value. This pattern is also present in  the \textbf{Entire loop} integration, with a smaller  contrast due to the average performed.
For both the \textbf{Central pixel} and the \textbf{Entire loop} integration, we see on the right panels of  Fig.~\ref{fig.deltaIbasem_freqs} and \ref{fig.deltaIbasem_whole_freqs}, respectively, that larger initial energy perturbations imply larger MDs. The blue, orange, green and red lines represent the B1, B3, B10 and B20 cases, respectively.

\subsection{Viewing angles analysis}

\begin{figure}[h!]\centering
\includegraphics[width=.48\linewidth]{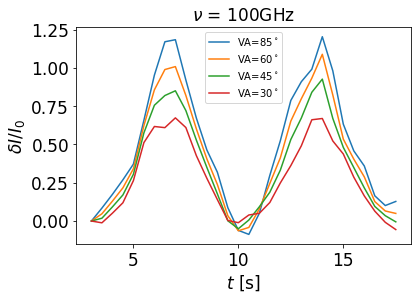}
\includegraphics[width=.48\linewidth]{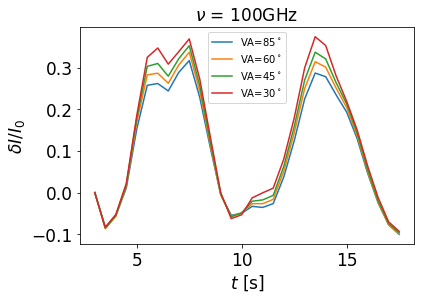}
\caption{MD analysis for different angles. Left panel: MD of the \textbf{Central pixel} integration for the B20 case. Right panel: same as left panel for the \textbf{Entire loop} integration.}
\label{fig.base20an}
\end{figure}

\begin{figure}[h!]\centering
\includegraphics[width=.75\linewidth]{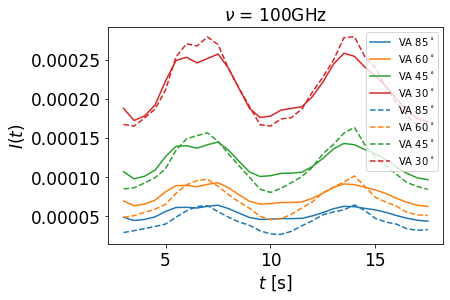}
\caption{Intensity of the \textbf{Central pixel} (dashed line) and the \textbf{Entire loop} (solid line) integration for the B20 case at different angles.}
\label{fig.wholeandcentral}
\end{figure}

Concerning different angles, we analyse the temporal MD for the \textbf{Central pixel} integration (left panel of Fig.~\ref{fig.base20an}), considering  $30^{\circ}$ (red line), $45^{\circ}$ (green line), $60^{\circ}$ (orange line) and $85^{\circ}$ (blue line) viewing  angles for the B20 case. Smaller MDs are obtained for smaller angles at $100~\text{GHz}$ (in the average the $30^{\circ}$ case is smaller in a  $43\%$ than $85^{\circ}$).
Considering the \textbf{Entire loop} integration, the MD is smaller than in the  \textbf{Central pixel} integration due to the average performed. However, the correlation between the MD and the variation of the LOS angle is opposite, i.e., the MD increases when the angle decreases (in the average the $30^{\circ}$ case is larger in a  $13\%$ than the $85^{\circ}$ case) (see right panel of the figure). From Fig.~\ref{fig.wholeandcentral}, we note as the angle is increased, for the {\bf Entire loop}, the average over the whole loop produces a flatter modulation in intensity, resulting in a smaller MD. Unlike R14, both the \textbf{Central pixel} integration and the \textbf{Entire loop} integration do not present relative phase changes, and the same occurs for the other frequencies ($10~\text{GHz}$ and $35~\text{GHz}$), not shown here.

\subsection{Mean modulation depth}

To quantify the intensity deviation from its average value we define the mean modulation depth (MMD) as $\Delta = 2 \max\left( \frac{I(t) - \langle I \rangle}{\langle I \rangle}\right)$, where $\langle I \rangle$ is the intensity average over a period.

\begin{figure*}[ht!]
\includegraphics[width=.33\linewidth]{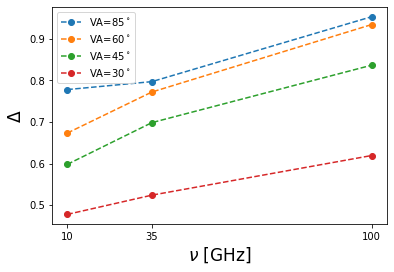}
\includegraphics[width=.33\linewidth]{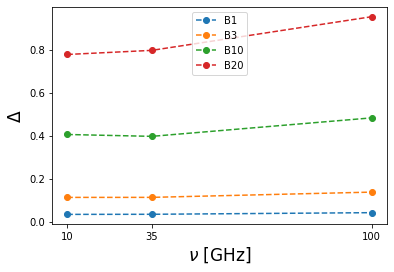}
\includegraphics[width=.33\linewidth]{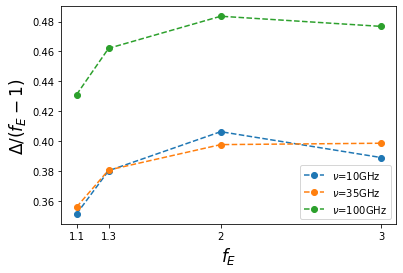}
\caption{{\bf Central pixel} MMD analysis. Left panel: MMD with frequency for case B20 at different viewing angles. Middle panel: Same as left panel for viewing angle $85^{\circ}$ at different cases. Right panel: Ratio between MMD and linear growth of energy deposition.}
\label{fig.md}
\end{figure*}

In Fig.~\ref{fig.md} we show the results found for the {\bf Central pixel} integration. The left and middle panels show a clear trend of the increase in the MMD curve as the frequency and the viewing angle increase, respectively.
Note that only the B20 case, with the larger viewing angle, reaches almost $100\%$ of the modulation depth, for the highest frequency.
As \cite{2016SoPh..291.3427K} we found that  our perturbations have finite amplitudes. Thus, the perturbations could excite sausage modes with finite amplitudes as well,  justifying  the deep MMD observed. This implies that we can use the high modulation depth cases as seismological tools to detect strong magnetic field cases.

We also compare the increment of the MMD with the increment of the energy deposition, i.e.  cases B1, B3, B10 and B20 have an increment of
$f_E=\Delta E_i/E_{i,\text{eq}} = (1.1, 1.3, 2, 3)$ with respect to  the equilibrium value, respectively. Thus,   we calculate the ratio between the MMD and the linear increment of the energy deposition centred at $0$ (to have a common origin for comparison), i.e. $\Delta/(f_E-1)$. From Fig.~\ref{fig.md} we note that, for all frequencies, the increase in MMD is not linear with the energy deposition increment. In fact, we note a saturation of the MMD increment with the energy deposition.

\begin{figure*}[ht!]
\includegraphics[width=.33\linewidth]{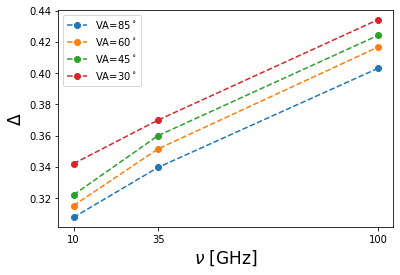}
\includegraphics[width=.33\linewidth]{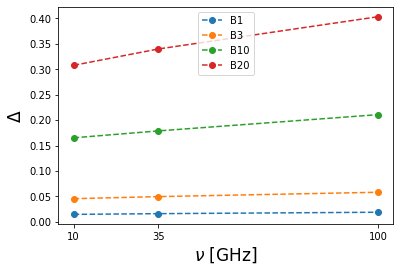}
\includegraphics[width=.33\linewidth]{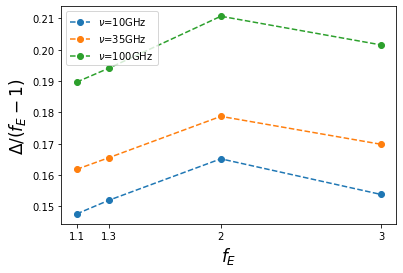}
\caption{Same as Fig.~\ref{fig.md} for the {\bf Entire loop}.}
\label{fig.md.el}
\end{figure*}

We perform the same analysis for the {\bf Entire loop} (Fig.~\ref{fig.md.el}). As a difference with the {\bf Central pixel} case the left panel clearly shows that, for a given frequency, the MMD increases when the viewing angles decrease. The middle and right panel show similar results as the {\bf Central pixel} case but with smaller values.

\section{Conclusions} \label{sec:conc}

Via  2.5D numerical simulations, and impulsively  perturbing a straight loop system (length of $25\,\text{Mm}$, and width of $6\,\text{Mm}$) of coronal parameters in equilibrium with its environment (chromosphere and corona),  we analysed the excitation of dominant fast sausage modes. A wavelet analysis performed over the GS radiation emission of a  \textbf{Central pixel} showed that, for all the energy perturbations used, a fast fundamental sausage mode of $\sim7\,\text{s}$ developed during $\sim170\,\text{s}$. Also, we found a first harmonic of the fast sausage mode of $3~\text{s}$ lasting for $\sim25~\text{s}$,  and a weak slow  sausage mode, below a wavelet confidence level.
The general behaviour of the radiation intensity obtained from   the wavelet analysis resembles the observational description shown in \citet{2020STP.....6a...3K} (see figure 5 of these authors), where the presence of a fast mode,  with its  first harmonic together with a similar decay pattern, suggests a common explanation of the phenomenon.

Our work shows that most of the results obtained by R14, where the excitation of the sausage mode was performed by a forced, monochromatic and linear perturbation, remain when an impulsive excitation of the mode
takes place,
i.e., when a typical fluctuation of the plasma parameters is able to impulsively trigger the oscillation and its corresponding
GS emission. As in R14, the increase of radio
brightness corresponds to the areas of stronger magnetic field
strength and higher density for all frequencies. Also, we found that for a  \textbf{Central pixel} integration, the larger the radio frequency and the larger the LOS angle, with respect to the loop axis, the larger the MD obtained.

We also considered the integration of the intensity over the \textbf{Entire loop}.
As in the \textbf{Central pixel} integration we found that the larger the frequency, the larger the MD obtained.  However,  we here found that the  larger the LOS angle, with respect to the loop axis, the smaller the MD obtained. This should be  a difference between high resolution  observations of sausage modes, as in solar loops, and measurement of sausage modes in low resolution, as in stellar flares observations. Thus, this eventually provides a tool to distinguish between sausage modes and other sources of QPP signals.

On the other hand, the increase of the energy deposition implies, for both spatial resolutions, an increase of the MD. For the \textbf{Central pixel} resolution and for the lower energy B1 case, considering $100~\text{GHz}$, the MD is  $\sim 104\%$,
for B3 is $\sim 113\%$, for B10 is $\sim 154\%$ and for B20 is $\sim 220\%$, with respect to their initial values, respectively. This could show that fast MHD modes triggered by impulsive energy depositions of the order of a microflare energy are able to reproduce deep  MDs in radio emission as occur in solar observations.
When considering the low resolution integration (the \textbf{Entire loop} integration) the increase of the MD is at the most of $132\%$ in the more energetic case of  B20.
The MD provides information on the deviation of intensity from its initial value, while the MMD describes the deviation of intensity from its mean value. This is important for the interpretation of observations. The MMD results show the same trends as the MD results. Also, in both cases, {\bf Central pixel} and {\bf Entire loop}, we found that the increment of the MMD is not linear with the increment of the energy deposition.

We schematically  summarise our results as: 1) We confirm that, for all spatial resolutions,   the larger the GS frequency of sausage modes, the larger the MD and MMD; 2) Depending on the high and low spatial resolution of the integration,  larger  LOS angles give rise to larger  MDs and MMDs  and smaller  MDs and MMDs, respectively; 3) We show that large MDs and MMD  are also associated with sausage modes when they are obtained considering more general contexts and perturbations; 4) The regularities found here suggest a way of differentiating sausage mode QPPs from QPPs of other origin; 5) These results would also allow to identify sausage mode signals in stellar flares.


\begin{acknowledgements}
     MC and AC are members of the Carrera del Investigador Cient\'ifico (CONICET). MC and AC acknowledge support from SECYT-UNC grant number PC No. 33620180101147CB and from PIP under grant number No. 11220200103150CO. TVD was supported by the European Research Council (ERC) under the European Union's Horizon 2020 research and innovation programme (grant agreement No 724326), the C1 grant TRACEspace of Internal Funds KU Leuven, and a Senior Research Project (G088021N) of the FWO Vlaanderen. Also, we thank the Centro de C\'omputo de Alto Desempe\~no (UNC), where the simulations were carried out. This research was supported by the International Space Science Institute (ISSI) in Bern, through ISSI International Team project N$^\circ$ 527 Bridging New X-ray Observations and Advanced Models of Flare Variability: A Key to Understanding the Fundamentals of Flare Energy Release.
\end{acknowledgements}

%
   \bibliographystyle{aa} 
   \bibliography{biblio} 
%

\end{document}